\documentstyle[12pt]{article}
\begin{document}

\begin{flushright}
{\bf hep-ph/9904360} \\
{\bf LMU-99-07} \\
{\bf PAR-LPTHE-99-17}\\
{April 1999}
\end{flushright}

\vspace{0.2cm}

\begin{center}
{\large\bf $CP$ Asymmetries in $B_d \rightarrow D^{*+}D^{*-}$ 
and $B_s\rightarrow D^{*+}_s D^{*-}_s$ Decays: \\
$P$-wave Dilution, Penguin and Rescattering Effects}
\end{center}

\vspace{.5cm}
\begin{center}
{\bf Xuan-Yem Pham} \footnote{
Postal address: LPTHE tour 16/1$^{\rm er}$ $\rm\acute{e}$tage,
Universit$\rm\acute{e}$ Pierre et Marie Curie, BP 126,
4 Place Jussieu, F-75252 Paris CEDEX 05, France;
Electronic address: pham@lpthe.jussieu.fr}\\
{\sl Laboratoire de Physique Th$\sl\acute{e}$orique et
Hautes Energies, Universit$\sl\acute{e}$s Pierre et
Marie Curie \\
(Paris 6) et Denis Diderot (Paris 7), 
Unit$\sl\acute{e}$ associ$\sl\acute{e}$e au CNRS, UMR 7589,
France}
\end{center} 

\begin{center}
{\bf Zhi-zhong Xing} \footnote{
Electronic address: xing$@$hep.physik.uni-muenchen.de }\\
{\sl Sektion Physik, Universit${\sl\ddot a}$t M${\sl\ddot
u}$nchen, Theresienstrasse 37A, D-80333 M${\sl\ddot u}$nchen, Germany}
\end{center}

\vspace{2.5cm}

\begin{abstract}
Determination of
the $CP$-violating
parameters $\sin 2\beta$ and $\sin 2\beta'$ is shown to be
possible from $B^0_d$ vs $\bar{B}^0_d\rightarrow D^{*+}D^{*-}$ 
and $B^0_s$ vs $\bar{B}^0_s\rightarrow D^{*+}_sD^{*-}_s$
decays {\it without} doing the angular analysis.
The $P$-wave dilution factors of these two asymmetries
are found to be 0.89 and 0.90, respectively,
using the factorization approximation and heavy quark symmetry.
The penguin-induced corrections amount to about $2\%$ and
$3\%$ in the corresponding $B_d$ and $B_s$ channels.  
Final-state rescattering effects could be handled by detecting
the neutral modes  $B^0_d$ and $\bar{B}^0_d\rightarrow
D^{*0}\bar{D}^{*0}$.
\end{abstract}

\newpage

{\Large\bf 1} ~
The study of $B$-meson decays appears to offer a unique 
opportunity to measure the quark mixing parameters,
to investigate the nonperturbative confinement forces,
and in particular to probe the origin of $CP$ violation.
Recently the CDF Collaboration \cite{CDF} has reported 
an updated direct measurement of the $CP$-violating parameter
$\sin 2\beta$  in $B^0_d$ vs $\bar{B}^0_d
\rightarrow J/\psi K_{\rm S}$ decays, 
where 
\begin{equation}
\beta \; = \; \arg \left ( - \frac{V^*_{tb}V_{td}}
{V^*_{cb}V_{cd}} \right ) \;\; 
\end{equation}
is known as 
an inner angle of the Cabibbo-Kobayashi-Maskawa (CKM) unitarity triangle.
The preliminary result 
$\sin2\beta = 0.79^{+0.41}_{-0.44} ~ ({\rm stat + syst})$
is consistent  with the standard-model prediction. If the CDF 
measurement 
is confirmed, $CP$ violation of the magnitude $\sin 2\beta$
should also be seen in 
$B^0_d$ vs $\bar{B}^0_d \rightarrow D^+D^-$, $D^{*+}D^-$,
$D^+D^{*-}$ and $D^{*+}D^{*-}$ decays \cite{Xing97,BB}, 
whose branching ratios are
all expected to be of order $10^{-4}$. Indeed the decay channel
$B^0_d \rightarrow D^{*+}D^{*-}$ has just been observed 
by the CLEO Collaboration \cite{CLEO}. The measured 
branching ratio 
${\cal B} (D^{*+}D^{*-}) = 
[ 6.2^{+4.0}_{-2.9} ~ ({\rm stat}) \pm 1.0 ~ ({\rm syst})]
\times 10^{-4}$
is in agreement with the standard-model
expectation. Further measurements of such 
decay modes will soon be
available in the first-round experiments of KEK and SLAC
$B$-meson factories as well as at other high-luminosity hadron
machines.

A comparison between the value of $\sin 2\beta$ 
to be determined from $B_d\rightarrow D^{*+}D^{*-}$ and that
already measured in 
$B_d\rightarrow J/\psi K_{\rm S}$ is no doubt important, as it 
may cross-check the consistency of the standard-model predictions.
Towards this goal, a
special attention has to be paid to possible uncertainties
associated with the $CP$ asymmetry in $B_d\rightarrow D^{*+}D^{*-}$. 
One kind of uncertainty comes from the penguin contamination,
as the weak phase of the penguin amplitude is 
quite different from that of the tree-level amplitude.
In contrast, the $CP$ asymmetry in the ``gold-plated'' modes  
$B^0_d$ vs $\bar{B}^0_d\rightarrow J/\psi K_{\rm S}$ is
essentially free from the penguin-induced 
uncertainty, since the relevant tree-level and penguin 
amplitudes almost have the same CKM phases.
Another 
kind of uncertainty arises from the $P$-wave dilution,
because the final state $D^{*+}D^{*-}$ is composed
of both the $CP$-even ($S$- and $D$-wave) 
and the $CP$-odd ($P$-wave) configurations. Of course an
analysis of the angular distributions of $B^0_d$ vs
$\bar{B}^0_d\rightarrow D^{*+}D^{*-}$ transitions allows us
to distinguish between the $CP$-even and $CP$-odd 
contributions \cite{AA}. In this work we point out that
the direct measurement of  $\beta$ can be made in 
$B_d\rightarrow D^{*+}D^{*-}$ decays without doing the angular analysis.  
Taking the $P$-wave dilution and the penguin contamination
into account, one may write
the characteristic measurable of indirect $CP$
violation in $B_d\rightarrow D^{*+}D^{*-}$ as follows
\footnote{The $CP$ violation induced solely by $B^0_d$-$\bar{B}^0_d$
mixing, like $\epsilon^{~}_K$ in the $K^0$-$\bar{K}^0$ mixing
system, is expected to be negligibly
small (of order $10^{-3}$ or smaller \cite{XingS97}) in the
standard model.}:
\begin{eqnarray}
\Delta_d & = & {\rm Im} \left (
\frac{V^*_{tb}V_{td}}{V_{tb}V^*_{td}} \cdot
\frac{\langle D^{*+}D^{*-}|{\cal H}_{\rm eff}|
\bar{B}^0_d\rangle}
{\langle D^{*+}D^{*-}|{\cal H}_{\rm eff}|B^0_d\rangle} \right ) 
\nonumber \\ \nonumber \\
& = & \zeta_d \left ( 1 - \xi_d \right )  \sin 2\beta \; ,
\end{eqnarray}
where $\zeta_d$ and $\xi_d$ 
represent the $P$-wave dilution factor and the penguin-induced
correction, respectively. We shall calculate both effects with
the help of the effective weak Hamiltonian and the factorization
approximation. Possible final-state rescattering effects in
$B_d\rightarrow D^{*+}D^{*-}$ decays will also be discussed.

A similar analysis will be made for $CP$ violation in the
decay modes $B^0_s$ vs $\bar{B}^0_s\rightarrow D^{*+}_sD^{*-}_s$.
The preliminary signals of these transitions
have been observed \cite{BS}. They are useful to
extract the $CP$-violating phase
\begin{equation}
\beta' \; = \; \arg \left ( - \frac{V^*_{tb}V_{ts}}
{V^*_{cb}V_{cs}} \right ) \;\; , 
\end{equation}
whose magnitude is negligibly small (of order $1^{\circ}$ or
smaller \cite{Xing98}) in the standard model. 
The associated $CP$ asymmetry
$\Delta_s$, analogous to $\Delta_d$ defined in Eq. (2), 
reads as 
\begin{eqnarray}
\Delta_s & = & {\rm Im} \left (
\frac{V^*_{tb}V_{ts}}{V_{tb}V^*_{ts}} \cdot
\frac{\langle D^{*+}_sD^{*-}_s|{\cal H}_{\rm eff}|
\bar{B}^0_s\rangle}
{\langle D^{*+}_sD^{*-}_s|{\cal H}_{\rm eff}|B^0_s\rangle} \right ) 
\nonumber \\ \nonumber \\
& = & \zeta_s \left ( 1 - \xi_s \right )  \sin 2\beta' \; ,
\end{eqnarray}
where $\zeta_s$ and $\xi_s$ denote 
the $P$-wave dilution factor and the penguin-induced
correction, respectively. Under SU(3) invariance $\zeta_s
= \zeta_d$ holds.  We shall see later on that 
$\zeta_d \approx \zeta_s \approx 0.9$, while the
magnitudes of $\xi_d$ and $\xi_s$ are only at the
percent level.

\vspace{0.5cm}

{\Large\bf 2} ~
Without loss of generality the amplitude of 
$B^0_d$ or $\bar{B}^0_d$ decay into $D^{*+}D^{*-}$ can be
written as a sum of three terms, i.e., the $S$- $D$- and
$P$-wave components \cite{Valencia}:
\begin{eqnarray}
\langle D^{*+}D^{*-}|{\cal H}_{\rm eff}|B^0_d\rangle
& = & a ~ (\epsilon_+ \cdot \epsilon_-) ~ + ~ \frac{b}{m^2_{D^*}}
~ (p^{~}_0 \cdot \epsilon_+) (p^{~}_0 \cdot \epsilon_-)
\nonumber \\
&  & + ~ i \frac{c}{m^2_{D^*}} ~ (\epsilon^{\alpha\beta\gamma\delta}
\epsilon_{+\alpha} \epsilon_{-\beta} p^{~}_{+\gamma}
p^{~}_{0\delta}) \; , \nonumber \\
\langle D^{*+}D^{*-}|{\cal H}_{\rm eff}|\bar{B}^0_d\rangle
& = & {\bar a} ~ (\epsilon_+ \cdot \epsilon_-) ~ + ~ \frac{\bar b}{m^2_{D^*}}
~ (p^{~}_0 \cdot \epsilon_+) (p^{~}_0 \cdot \epsilon_-)
\nonumber \\
&  & - ~ i \frac{\bar c}{m^2_{D^*}} ~ (\epsilon^{\alpha\beta\gamma\delta}
\epsilon_{+\alpha} \epsilon_{-\beta} p^{~}_{+\gamma}
p^{~}_{0\delta}) \; , 
\end{eqnarray}
where $\epsilon_{\pm}$ denotes the polarization of $D^{*\pm}$;
$p^{~}_0$ and $p^{~}_{\pm}$ stand respectively for the momenta of $B_d$ and
$D^{*\pm}$ mesons; $(a,b,c)$ and $(\bar{a},\bar{b},\bar{c})$ are
complex scalars. To calculate these scalars we neglect effects from the
annihilation-type quark diagrams which are anticipated to have
significant form-factor suppression. In this case
the effective weak Hamiltonian responsible for $B_d\rightarrow
D^{*+}D^{*-}$ decays can simply be written as \cite{Buras}
\footnote{Here we assume the top-quark dominance in the penguin loops
and neglect the small strong phases induced by absorptive parts of the
up and charm penguin loop-integral functions \cite{Penguin}.}
\begin{equation}
{\cal H}_{\rm eff} \; =\; \frac{G_{\rm F}}{\sqrt{2}}
\left [ (V_{cb}V^*_{cd}) \sum^2_{i=1} (c_i Q^c_i) 
- (V_{tb}V^*_{td}) \sum^{10}_{i=3} (c_i Q_i) \right ] 
~ + ~ {\rm h.c.} \; ,
\end{equation}
where $c_i$ (for $i=1, \cdot\cdot\cdot, 10$)
are the Wilson coefficients, and
\begin{eqnarray}
Q^c_1 & = & (\bar{d}_\alpha c_\beta )^{~}_{\rm V-A}
(\bar{c}_\beta b_\alpha )^{~}_{\rm V-A} \; ,
\nonumber \\
Q^c_2 & = & (\bar{d}c )^{~}_{\rm V-A} 
(\bar{c}b )^{~}_{\rm V-A} \; , 
\nonumber \\
Q_3 & = & (\bar{d}b )^{~}_{\rm V-A} 
(\bar{c}c )^{~}_{\rm V-A} \; ,
\nonumber \\
Q_4 & = & (\bar{d}_\alpha b_\beta )^{~}_{\rm V-A}
(\bar{c}_\beta c_\alpha )^{~}_{\rm V-A} \; ,
\nonumber \\
Q_5 & = & (\bar{d}b )^{~}_{\rm V-A} 
(\bar{c}c )^{~}_{\rm V+A} \; ,
\nonumber \\
Q_6 & = & (\bar{d}_\alpha b_\beta )^{~}_{\rm V-A}
(\bar{c}_\beta c_\alpha )^{~}_{\rm V+A} \; ,
\end{eqnarray}
as well as $Q_7 = Q_5$, $Q_8 = Q_6$, $Q_9 = Q_3$ and $Q_{10} = Q_4$. 
Here $Q_3, \cdot\cdot\cdot ,Q_6$ denote the QCD-induced penguin
operators, and $Q_7, \cdot\cdot\cdot ,Q_{10}$ stand for the electroweak
penguin operators. The factorization approximation allows us to
single out a common hadronic matrix element from
$\langle D^{*+}D^{*-}|{\cal H}_{\rm eff}|B^0_d\rangle$
and another one from
$\langle D^{*+}D^{*-}|{\cal H}_{\rm eff}|\bar{B}^0_d\rangle$:
\begin{eqnarray}
M & = & \frac{G_{\rm F}}{\sqrt{2}} 
\langle D^{*+}|(\bar{c}d)^{~}_{\rm V-A}|0\rangle
\langle D^{*-}|(\bar{b}c)^{~}_{\rm V-A}|B^0_d\rangle \; ,
\nonumber \\
\bar{M} & = & \frac{G_{\rm F}}{\sqrt{2}} 
\langle D^{*-}|(\bar{d}c)^{~}_{\rm V-A}
|0\rangle \langle D^{*+}|(\bar{c}b)^{~}_{\rm V-A}|\bar{B}^0_d\rangle \; .
\end{eqnarray}
In this approach it should be noted that the relevant
Wilson coefficients
and the hadronic matrix elements need be evaluated in the same
renormalization scheme and at the same energy scale \cite{Buras}.
Furthermore, we follow the standard procedure of Ref. \cite{BSW} to
decompose $M$ and $\bar{M}$ in terms of three form factors 
$A^{BD^*}_1(q^2)$,
$A^{BD^*}_2(q^2)$ and $V^{BD^*}(q^2)$. 
Then the scalars $(a,b,c)$ and $(\bar{a}, \bar{b}, \bar{c})$ 
are found to be
\begin{eqnarray}
(a, ~ b, ~ c ) & = & \frac{G_{\rm F}}{\sqrt{2}} ~ \chi ~
(\tilde{a}, ~ \tilde{b}, ~ \tilde{c}) \; , 
\nonumber \\ 
(\bar{a}, ~ \bar{b}, ~ \bar{c}) & = & \frac{G_{\rm F}}{\sqrt{2}} ~ \chi^*
(\tilde{a}, ~ \tilde{b}, ~ \tilde{c}) \; , 
\end{eqnarray}
where
\begin{equation}
\chi \; =\; (V^*_{cb}V_{cd}) c_x - (V^*_{tb}V_{td})
(c_y + c_z) \; 
\end{equation}
with
\begin{eqnarray}
c_x & = & \bar{c}_2 + \frac{\bar{c}_1}{3} \; ,
\nonumber \\ \nonumber \\
c_y & = & \bar{c}_4 + \frac{\bar{c}_3}{3} \; ,
\nonumber \\ \nonumber \\
c_z & = & \bar{c}_{10} + \frac{\bar{c}_9}{3} \; ;
\end{eqnarray}
and
\begin{eqnarray}
\tilde{a} & = & m^{~}_{D^*} f_{D^*} (m^{~}_B + m^{~}_{D^*})
A^{BD^*}_1(m^2_{D^*}) \; ,
\nonumber \\ \nonumber \\
\tilde{b} & = & -2 m^3_{D^*} f_{D^*} ~ \frac{A^{BD^*}_2(m^2_{D^*})}
{m^{~}_B + m^{~}_{D^*}} \; ,
\nonumber \\ \nonumber \\
\tilde{c} & = & -2 m^3_{D^*} f_{D^*} ~ \frac{V^{BD^*}(m^2_{D^*})}
{m^{~}_B + m^{~}_{D^*}} \; . 
\end{eqnarray}
Note that in Eq. (11) the effective
Wilson coefficients $\bar{c}_i$ are independent of the energy
scale and the renormalization scheme \cite{Buras}.
It is clear that
$c_x$, $c_y$ and $c_z$ stand for the tree-level, gluonic penguin
and electroweak penguin contributions, respectively. 
Taking the top-quark mass $m_t = 174$ GeV and the strong
coupling constant $\alpha_s(m_b) = 0.21$, one finds
$c_x = 1.045$, $c_y =-0.031$ and $c_z =-0.0014$ \cite{He}.
Note also 
that the form factors in Eq. (12) are related to one another in the 
heavy quark symmetry \cite{Pham98}. The $q^2$ dependence of the form 
factors, given by the common slope of the universal 
Isgur--Wise function, allows us to extrapolate from $q^2_{\rm max} = 
(m^{~}_B - m^{~}_{D^*})^2$ to $q^2 = m^2_{D^*}$. Then we get
\begin{equation}
V^{BD^*}(m^2_{D^*}) \; =\; A^{BD^*}_2(m^2_{D^*}) 
\; =\; \frac{(m^{~}_B + m^{~}_{D^*})^2}{
 m^{~}_B(m^{~}_B + 2 m^{~}_{D^*})} A_1^{BD^*}(m^2_{D^*})  \; .
\end{equation}
Therefore the ratios $\tilde{b}/\tilde{a} = \tilde{c}/\tilde{a}$ 
depend only upon the meson masses $m^{~}_B$ and $m^{~}_{D^*}$.

Now let us determine the $\zeta_d$ and $\xi_d$ parameters in 
the $CP$-violating quantity $\Delta_d$ defined in Eq. (2). For this
purpose we sum over the polarizations of $D^{*+}$ and
$D^{*-}$ mesons \cite{Valencia}. 
After a lengthy but straightforward
calculation, we arrive at
\begin{eqnarray}
&  & \sum_{\rm (pol)} \left (
\frac{\langle D^{*+}D^{*-}|{\cal H}_{\rm eff}|\bar{B}^0_d\rangle}
{\langle D^{*+}D^{*-}|{\cal H}_{\rm eff}|B^0_d\rangle}
\right ) 
\nonumber \\ \nonumber \\
& = & \frac{(2+\kappa^2) (\bar{a}a^*) + (\kappa^2-1)^2 (\bar{b}b^*)
+ \kappa (\kappa^2-1) (\bar{a}b^* + \bar{b} a^*)
- 2(\kappa^2-1)(\bar{c}c^*)}
{(2+ \kappa^2) |a|^2 + (\kappa^2 -1)^2 |b|^2 
+ 2 \kappa (\kappa^2 -1) {\rm Re} (ab^*)
+ 2 (\kappa^2 -1) |c|^2} 
\nonumber \\ \nonumber \\
& = & \frac{\chi^*}{\chi} \cdot 
\frac{(2+\kappa^2) \tilde{a}^2 + (\kappa^2 -1)^2 \tilde{b}^2
+ 2 \kappa (\kappa^2 -1) \tilde{a} \tilde{b}
- 2 (\kappa^2 -1) \tilde{c}^2}
{(2+\kappa^2) \tilde{a}^2 + (\kappa^2 -1)^2 \tilde{b}^2
+ 2 \kappa (\kappa^2 -1) \tilde{a} \tilde{b} 
+ 2 (\kappa^2 -1) \tilde{c}^2}
\; ,
\end{eqnarray}
where 
$\kappa = (p_+\cdot p_-)/m^2_{D^*} =(m^2_B-2m^2_{D^*})/(2m^2_{D^*})$.
As a result, we obtain
\begin{equation}
\zeta_d\; =\; \frac{m^3_B - 3 m^{~}_B m^2_{D^*} + 10 m^3_{D^*}}
{m^3_B + m^{~}_B m^2_{D^*} + 2 m^3_{D^*}} \;\; ,
\end{equation}
and 
\begin{equation}
\xi_d \; =\; \frac{c_y + c_z}{c_x} \cdot \frac{\cos 2\beta}{\cos\beta}
\cdot \left | \frac{V_{tb}V_{td}}{V_{cb}V_{cd}} \right | 
\;\; .
\end{equation}
It is remarkable that the result obtained in Eq. (15) for the 
$P$-wave dilution factor $\zeta_d$ relies only on 
the heavy quark symmetry. Therefore the value of $\zeta_d$ 
is independent of specific models for the form factors. 
To estimate the penguin-induced correction (i.e., the $\xi_d$ parameter), 
$|(V_{tb}V_{td})/(V_{cb}V_{cd})|=1$ and $\beta = 26^{\circ}$ 
are typically taken. The former is favored by current
data on quark mixing \cite{PDG98}, and the latter
is consistent with
$\sin 2\beta =0.79$ observed by the CDF Collaboration \cite{CDF}.
Then we find
\begin{eqnarray}
\zeta_d & = & 0.89 \; , \nonumber \\
\xi_d & = & -0.021 \; .
\end{eqnarray}
This result indicates that the penguin contamination in $\Delta_d$
is negligibly small, while the $P$-wave dilution to $\Delta_d$
should be taken seriously. 

For $B^0_s$ vs $\bar{B}^0_s\rightarrow D^{*+}_s D^{*-}_s$ decays,
the $\zeta_s$ and $\xi_s$ parameters appearing in the $CP$ asymmetry
$\Delta_s$ can be evaluated in the same way. The relevant results
are obtained, through the discrete transformation from $d$ to
$s$ (the so-called U-spin reflection) in the
above formulas, as follows:
\begin{equation}
\zeta_s\; =\; \frac{m^3_{B_s} - 3 m^{~}_{B_s} m^2_{D^*_s} + 10 m^3_{D^*_s}}
{m^3_{B_s} + m^{~}_{B_s} m^2_{D^*_s} + 2 m^3_{D^*_s}} \;\; ,
\end{equation}
and 
\begin{equation}
\xi_s \; =\; \frac{c_y + c_z}{c_x} \cdot \frac{\cos 2\beta'}{\cos\beta'}
\cdot \left | \frac{V_{tb}V_{ts}}{V_{cb}V_{cs}} \right | 
\;\; .
\end{equation}
For illustration, we typically take $|(V_{tb}V_{ts})/(V_{cb}V_{cs})| =1$ 
and $\beta' =0.5^{\circ}$ to calculate the value of $\xi_s$.
We obtain 
\begin{eqnarray}
\zeta_s & = & 0.90 \; , \nonumber \\
\xi_s & = & -0.031 \; .
\end{eqnarray}
Indeed $\zeta_s = \zeta_d$ is expected to hold exactly under SU(3) symmetry.
The result in Eq. (20), similar to that in Eq. (17) for $B_d\rightarrow
D^{*+}D^{*-}$ modes, implies that in $\Delta_s$ the penguin contamination
is negligibly small but the $P$-wave dilution is significant. 

\vspace{0.5cm}

{\Large\bf 3} ~
Now we proceed to discuss possible final-state rescattering effects in
$B_d\rightarrow D^{*+}D^{*-}$ decays, which were not taken into
account in the above analysis. 
The $\Delta B =+1$ and $\Delta B =-1$ parts of 
${\cal H}_{\rm eff}$ in Eq. (6) have the isospin 
structures $|1/2, +1/2\rangle$ and $|1/2, -1/2\rangle$,
respectively. They generally govern the transitions
$B^+_u\rightarrow D^{*+}\bar{D}^{*0}$, $B^0_d\rightarrow
D^{*+}D^{*-}$, $B^0_d\rightarrow D^{*0}
\bar{D}^{*0}$ and their charge-conjugate processes. The
final state of each decay mode 
can be in either $I=1$ or $I=0$ isospin configuration,
therefore rescattering effects are possibly present.
We find that the isospin relations \cite{Xing97,Pham87}
\begin{eqnarray}
\langle D^{*+}\bar{D}^{*0}|{\cal H}_{\rm eff}|B^+_u\rangle
& = & A_1 \; , \nonumber \\
\langle D^{*+}D^{*-}|{\cal H}_{\rm eff}|B^0_d\rangle 
& = & \frac{1}{2} (A_1 + A_0 ) \; , \nonumber \\
\langle D^{*0}\bar{D}^{*0}|{\cal H}_{\rm eff}|B^0_d\rangle 
& = & \frac{1}{2} (A_1 - A_0 ) \; ,
\end{eqnarray}
where $A_1$ and $A_0$ are the $I=1$ and $I=0$ isospin amplitudes,
hold separately for three transition amplitudes with the same
helicity ($\lambda =-1$, 0 or $+1$). The same isospin relations
can be obtained for $B^-_u\rightarrow D^{*-}D^{*0}$,
$\bar{B}^0_d\rightarrow D^{*+}D^{*-}$ and
$\bar{B}^0_d\rightarrow D^{*0}\bar{D}^{*0}$ decays in terms of
the corresponding isospin amplitudes $\bar{A}_1$ and $\bar{A}_0$.

To calculate the magnitudes of $A_1$, $A_0$ and $\bar{A}_1$ and
$\bar{A}_0$, we assume again that transition amplitudes from the 
annihilation-type quark diagrams are negligible
due to their strong form-factor suppression
\footnote{For example, hadronic matrix elements of the type
$\langle D^{*+}D^{*-}|(\bar{c}c)^{~}_{\rm V-A}|0\rangle
\langle 0|(\bar{b}d)^{~}_{\rm V-A}|B^0_d\rangle$ depend
on the annihilation form factor $F^{\rm ann}_+(m^2_B)
\sim 1/m^2_B$ \cite{Brodsky}
and the mass difference of two final-state
mesons in a constituent U(2,2) quark model
\cite{Xing96}.}.
This implies that $B^0_d\rightarrow D^{*0}\bar{D}^{*0}$ and
its charge-conjugate process, which occur only through
the quark diagrams illustrated in Fig. 1, 
\begin{figure}
\begin{picture}(400,200)(-20,50)
\put(70,240){\line(1,0){90}}
\put(70,190){\line(1,0){90}}
\put(62,236){$\bar{b}$}
\put(62,186){$d$}
\put(40,210.5){$B^0_d$}
\put(163,238){$\bar{c}$}
\put(163,187){$c$}
\put(163,225){$u$}
\put(163,200){$\bar{u}$}
\put(176,191){$D^{*0}$}
\put(176,230){$\bar{D}^{*0}$}
\put(160,215){\oval(70,25)[l]}
\put(85,240){\vector(-1,0){2}}
\put(85,190){\vector(1,0){2}}
\put(145,240){\vector(-1,0){2}}
\put(145,190){\vector(1,0){2}}
\put(145,227.5){\vector(1,0){2}}
\put(145,202.5){\vector(-1,0){2}}
\multiput(100,233.6)(0,-6.1){8}{$>$}
\put(115,168){(a)}
\put(280,240){\line(1,0){90}}
\put(280,190){\line(1,0){90}}
\put(272,236){$\bar{b}$}
\put(272,186){$d$}
\put(250,210.5){$B^0_d$}
\put(373,238){$\bar{u}$}
\put(373,187){$u$}
\put(373,225){$c$}
\put(373,200){$\bar{c}$}
\put(386,191){$\bar{D}^{*0}$}
\put(386,230){$D^{*0}$}
\put(370,215){\oval(70,25)[l]}
\put(295,240){\vector(-1,0){2}}
\put(295,190){\vector(1,0){2}}
\put(355,240){\vector(-1,0){2}}
\put(355,190){\vector(1,0){2}}
\put(355,227.5){\vector(1,0){2}}
\put(355,202.5){\vector(-1,0){2}}
\multiput(310,233.6)(0,-6.1){8}{$>$}
\put(325,168){($\rm a'$)}
\put(70,120){\oval(84,31)[r]}
\put(160,120){\oval(52,21)[l]}
\put(160,120){\oval(74,46)[l]}
\put(85,135.5){\vector(-1,0){2}}
\put(85,104.5){\vector(1,0){2}}
\put(145,143){\vector(-1,0){2}}
\put(145,97){\vector(1,0){2}}
\put(145,130.5){\vector(1,0){2}}
\put(145,109.5){\vector(-1,0){2}}
\multiput(92,129.5)(0,-6){5}{$>$}
\put(40,118){$B^0_d$}
\put(62,132){$\bar{b}$}
\put(62,102){$d$}
\put(163,141){$\bar{c}$}
\put(163,95){$c$}
\put(163,129){$u$}
\put(163,107){$\bar{u}$}
\put(176,134){$\bar{D}^{*0}$}
\put(176,99){$D^{*0}$}
\put(115,77){(b)}
\put(280,120){\oval(84,31)[r]}
\put(370,120){\oval(52,21)[l]}
\put(370,120){\oval(74,46)[l]}
\put(295,135.5){\vector(-1,0){2}}
\put(295,104.5){\vector(1,0){2}}
\put(355,143){\vector(-1,0){2}}
\put(355,97){\vector(1,0){2}}
\put(355,130.5){\vector(1,0){2}}
\put(355,109.5){\vector(-1,0){2}}
\multiput(302,129.5)(0,-6){5}{$>$}
\put(250,118){$B^0_d$}
\put(272,132){$\bar{b}$}
\put(272,102){$d$}
\put(373,141){$\bar{u}$}
\put(373,95){$u$}
\put(373,129){$c$}
\put(373,107){$\bar{c}$}
\put(386,134){$D^{*0}$}
\put(386,99){$\bar{D}^{*0}$}
\put(325,77){($\rm b'$)}
\end{picture}
\vspace{-0.6cm}
\caption{Tree-level (a, $\rm a'$) and penguin (b, $\rm b'$)
quark diagrams responsible for $B^0_d\rightarrow D^{*0}\bar{D}^{*0}$.}
\end{figure}
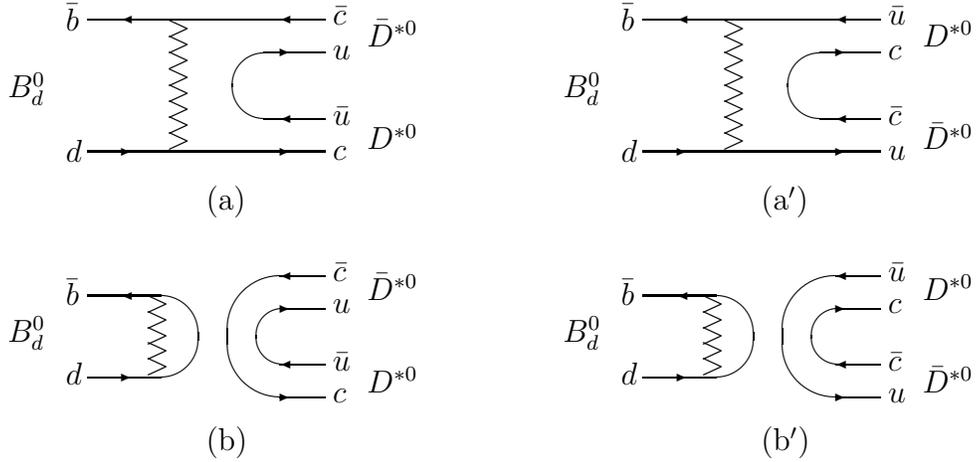
would be forbidden if
there were no final-state rescattering. In other words,
$A_1 = A_0$ and $\bar{A}_1 = \bar{A}_0$ would hold, 
if the rescattering effect were absent. Following this
argument we make use of the factorization approximation 
to calculate $A_1$ (or $\bar{A}_1$) and $A_0$ (or $\bar{A}_0$),
and account for final-state
interactions at the hadron level by incorporating the elastic
rescattering phases $\delta_1$ and $\delta_0$ \cite{Neubert}. 
We then arrive
at the factorized isospin amplitudes as follows:
\begin{eqnarray}
A_1 & = & \chi M e^{i\delta_1} \; , 
\nonumber \\
A_0 & = & \chi M e^{i\delta_0} \; ;
\end{eqnarray}
and
\begin{eqnarray}
\bar{A}_1 & = & \chi^* \bar{M} e^{i\delta_1} \; , 
\nonumber \\
\bar{A}_0 & = & \chi^* \bar{M} e^{i\delta_0} \; ,
\end{eqnarray}
where $M$ and $\bar{M}$ are given in Eq. (8).
One can see that $|A_0| = |A_1|$ and $|\bar{A}_0|=|\bar{A}_1|$
hold in this factorization approach.
It becomes clear that the transitions $B_d\rightarrow D^{*0} \bar{D}^{*0}$
would be forbidden, if there were no final-state
rescattering effects (i.e., if $\delta_0 = \delta_1$). 
As a consequence, one obtains
\begin{eqnarray}
\langle D^{*+}D^{*-}|{\cal H}_{\rm eff}|B^0_d\rangle 
& = & \chi ~ M \cos\frac{\delta_1 -\delta_0}{2} 
e^{i (\delta_1 + \delta_0)/2} \; , 
\nonumber \\
\langle D^{*+}D^{*-}|{\cal H}_{\rm eff}|\bar{B}^0_d\rangle
& = & \chi^* \bar{M} \cos\frac{\delta_1 -\delta_0}{2} 
e^{i (\delta_1 + \delta_0)/2} \; ;
\end{eqnarray}
and
\begin{eqnarray}
\langle D^{*0}\bar{D}^{*0}|{\cal H}_{\rm eff}|B^0_d\rangle 
& = & i \chi ~ M \sin\frac{\delta_1 -\delta_0}{2} 
e^{i (\delta_1 + \delta_0)/2} \; , 
\nonumber \\
\langle D^{*0}\bar{D}^{*0}|{\cal H}_{\rm eff}|\bar{B}^0_d\rangle
& = & i \chi^* \bar{M} \sin\frac{\delta_1 -\delta_0}{2} 
e^{i (\delta_1 + \delta_0)/2} \; .
\end{eqnarray}
We see that the isospin phases can be cancelled out in the ratio of 
two charge-conjugate decay amplitudes, thus the previous results of $\zeta_d$,
$\xi_d$ and $\Delta_d$ are unchanged even in the presence
of final-state rescattering effects. Of course this conclusion depends 
on the factorization hypothesis used above. 
Whether it is valid or not
can be checked, once the relevant branching ratios are 
measured. For example, Eqs. (24) and (25) lead to two
experimentally testable
relations among the branching ratios of three correlative
decay modes:
\begin{eqnarray}
|\langle D^{*+}D^{*-}|{\cal H}_{\rm eff}|B^0_d\rangle|^2
~ + ~ |\langle D^{*0}\bar{D}^{*0}|{\cal H}_{\rm eff}|B^0_d\rangle|^2
& = & |\langle D^{*+}\bar{D}^{*0}|{\cal H}_{\rm eff}|B^+_u\rangle|^2
\; , \nonumber \\
|\langle D^{*+}D^{*-}|{\cal H}_{\rm eff}|\bar{B}^0_d\rangle|^2
~ + ~ |\langle D^{*0}\bar{D}^{*0}|{\cal H}_{\rm eff}|\bar{B}^0_d\rangle|^2
& = & |\langle D^{*-}D^{*0}|{\cal H}_{\rm eff}|B^-_u\rangle|^2
\; .
\end{eqnarray}
It can easily be shown that $|M|^2 = |\bar{M}|^2$ will hold,
if one sums over the polarizations of two final-state vector mesons.
In this case, the two (rectangular) triangle relations in Eq. (26)
are congruent with each other.

It is worth remarking that the detection of
$B_d \rightarrow D^{*0}\bar{D}^{*0}$
transitions  will be crucial: (a) if their 
branching ratios in comparison with
those of $B_d \rightarrow D^{*+}D^{*-}$ are too small to
be observed, then the final-state rescattering
effects should be negligible ($\delta_1 -\delta_0 \approx 0$) 
and the naive factorization
approach might work well; (b) if their branching
ratios are more or less comparable with those of
$B_d \rightarrow D^{*+}D^{*-}$, then a quantitative
isospin analysis should be available, allowing 
us to extract the isospin phase differences through
Eq. (21). In case (b), a measurement of $CP$ violation of the
magnitude $\sin 2\beta$ in $B^0_d$ vs $\bar{B}^0_d\rightarrow
D^{*0}\bar{D}^{*0}$ decays should be quite likely.

For $B^0_s$ and $\bar{B}^0_s$ decays into the
$D^{*+}_sD^{*-}_s$ state the similar isospin analysis does not
exist. The SU(3) symmetry between $B_d\rightarrow
D^{*+}D^{*-}$ and $B_s\rightarrow D^{*+}_sD^{*-}_s$ 
channels, however, allows one to conjecture possible
final-state interactions in the latter. As one has seen
in Eqs. (15) and (18), the SU(3) breaking effect
is indeed rather small and even negligible.

\vspace{0.5cm}

{\Large\bf 4} ~
We have calculated the indirect $CP$ asymmetries 
($\Delta_d$ and $\Delta_s$) in $B_d\rightarrow D^{*+}D^{*-}$
and $B_s\rightarrow D^{*+}_s D^{*-}_s$ decay modes. 
It has been shown that a quite clean determination of
the $CP$-violating parameters $\sin 2\beta$ and 
$\sin 2\beta'$, with no help of the angular analysis, is
in practice possible. The penguin contamination in either
$\Delta_d$ or $\Delta_s$ is negligibly small. The $P$-wave
dilution factors of $\Delta_d$ and $\Delta_s$ (i.e.,
$\zeta_d$ and $\zeta_s$) are found to
be 0.89 and 0.90, respectively, in the factorization approximation
and heavy quark symmetry. 

As $\zeta_d$ does not deviate too much from unity, a large $CP$
asymmetry in $B^0_d$ vs $\bar{B}^0_d\rightarrow D^{*+}D^{*-}$
transitions is expected within the standard model.
If the penguin-induced correction to the indirect $CP$ violation
in $B^0_d$ vs $\bar{B}^0_d\rightarrow D^+D^-$ is also
negligible, then a comparison between the $CP$ asymmetries in
$D^+D^-$ and $D^{*+}D^{*-}$ modes allows us to directly
determine the $P$-wave dilution factor, i.e.,
\begin{equation}
\zeta_d \; =\; \frac{\Delta_d (D^{*+}D^{*-})}
{\Delta_d (D^+D^-)} \; .
\end{equation}
Of course such a measurement is very useful, in order to
check the theoretical value of $\zeta_d$ obtained in Eq. (15).

Although the $CP$ asymmetry $\Delta_s$ is expected to be
vanishingly small in the standard model (because of
the smallness of $\beta'$), its magnitude 
could significantly be enhanced if there were new physics in
$B^0_s$-$\bar{B}^0_s$ mixing. 
For illustration let us consider
a kind of new physics 
that does not violate unitarity of the
CKM matrix and has insignificant effects on the penguin channels
of the decay modes under discussion \cite{XingS97}. 
It may introduce an additional $CP$-violating phase
into $B^0_d$-$\bar{B}^0_d$ or $B^0_s$-$\bar{B}^0_s$ mixing. 
In this case the overall mixing phases of two systems
become
\begin{eqnarray}
\frac{V^*_{tb}V_{td}}{V_{tb}V^*_{td}}
& \Longrightarrow & \frac{V^*_{tb}V_{td}}{V_{tb}
V^*_{td}} ~ e^{{\rm i}\phi_{\rm NP}} \; , 
\nonumber \\ \nonumber \\
\frac{V^*_{tb}V_{ts}}{V_{tb}V^*_{ts}}
& \Longrightarrow & \frac{V^*_{tb}V_{ts}}
{V_{tb}V^*_{ts}} ~ e^{{\rm i}\phi'_{\rm NP}} \; ,
\end{eqnarray}
in which $\phi_{\rm NP}$ and $\phi'_{\rm NP}$ denote
the $CP$-violating phases induced by new physics. The 
weak phases that can be extracted from the $CP$ asymmetries
$\Delta_d$ and $\Delta_s$ turn out to be
\begin{eqnarray}
\beta & \Longrightarrow & \beta ~ + ~ \frac{\phi_{\rm NP}}{2} \; ,
\nonumber \\
\beta' & \Longrightarrow & \beta' ~ + ~ \frac{\phi'_{\rm NP}}{2} \; ,
\end{eqnarray}
respectively. To distinguish $\beta$ and $\beta'$ from the
phase combinations in Eq. (29), one has to study the $CP$
asymmetries in some other neutral $B$-meson decays \cite{Xing98}.

In conclusion, we point out that the $CP$-violating
parameters in $B_d\rightarrow D^{*+}D^{*-}$ and
$B_s\rightarrow D^{*+}_sD^{*-}_s$ decays can be determined
without measuring their angular distributions.
The approach advocated here may be complementary to the 
angular analysis considered in the literature. Hopefully both will soon be 
confronted with the data from $B$-meson factories.

\vspace{0.5cm}

\underline{Acknowledgment:} ~ 
We would like to thank B. Machet for interesting discussions.

\end{document}